\documentclass[prl,reprint,amssymb,preprint,superscriptaddress,showpacs]{revtex4}
\usepackage{graphicx}

\begin{document}

\title{Cation-ordering Effects in the Single Layered Manganite La$_{2/3}$Sr$_{4/3}$MnO$_4$}

\author{B. B. Nelson-Cheeseman}
\email{bbnelsonchee@anl.gov} \affiliation{Materials Science
Division, Argonne National Laboratory, Argonne, IL 60439}
\author{A. B. Shah}
\affiliation{Department of Materials Science and Engineering, The
University of Illinois, Urbana-Champaign, IL 61801}
\author{T. S. Santos}
\affiliation{Center for Nanoscale Materials, Argonne National
Laboratory, Argonne, IL 60439}
\author{S. D. Bader}
\affiliation{Materials Science
Division, Argonne National Laboratory, Argonne, IL 60439}
\author{J.-M. Zuo}
\affiliation{Department of Materials Science and Engineering, The
University of Illinois, Urbana-Champaign, IL 61801}
\author{A. Bhattacharya}
\affiliation{Materials Science
Division, Argonne National Laboratory, Argonne, IL 60439}\affiliation{Center for Nanoscale Materials, Argonne National
Laboratory, Argonne, IL 60439}

\date{\today}

\begin{abstract}
We have synthesized epitaxial La$_{1-x}$Sr$_{1+x}$MnO$_4$ (x=1/3) films as random alloys and cation-ordered analogues to probe how cation order affects the properties of a 2D manganite. The films show weak ferromagnetic ordering up to 130 K, although there is a dramatic difference in magnetic anisotropy depending on the cation order. While all films exhibit similar gapped insulator behavior above 130 K, there is a significant difference in the low temperature transport mechanism depending on the cation order. Differences in magnetic anisotropy and low temperature transport are consistent with differences in Mn 3d orbital occupancies. Together this work suggests that cation ordering can significantly alter the Mn 3d orbital ground state in these correlated electron systems.\end{abstract}


\maketitle
The 3-dimensional and layered La,Sr-doped manganites, (La$_{1-x}$Sr$_{x}$MnO$_{3}$)$_{n}$SrO, exhibit a delicate interplay between charge-, spin- and orbital- degrees of freedom which gives rise to a number of intriguing collective states. Depending on cation doping and dimensionality, these materials can exhibit ferromagnetic-, antiferromagnetic-, charge- and orbital-ordering.\cite{Kimura00} While the A-site cation doping level is pivotal in determining the filling of the Mn e$_g$ orbitals, the dimensionality of the Mn octahedral network also plays a key role as these collective states may be highly susceptible to fluctuations and disorder that thwart long range order in lower dimensions.

In this letter, we address the effect of cation disorder in a 2-dimensional (2D)material which lacks the ordered state found at the same dopant level in higher dimensions. For a doping level of x=1/3 (2Mn$^{3+}$:Mn$^{4+}$), reducing the dimensionality of the Mn-octahedral network to isolated 2D sheets results in a rapid degradation of the robust ferromagnetism seen in the 3-dimensional (3D) manganite, La$_{2/3}$Sr$_{1/3}$MnO$_3$.  While long-range ferromagnetic order and metallic conductivity occurs in the bilayer and 3D compounds (T$_C$=120K and 360K, respectively), the equivalent single layer manganite exhibits only a spin glass phase below 20 K with no long-range magnetic order at any temperature and is an insulator.\cite{Konishi98,Larochelle05,Moritomo95} X-ray studies show that short range structural order associated with the charge-ordered state does appear in the guise of small correlated nanopatches; however, perpendicular to the MnO$_{2}$ planes the material never obtains full magnetic long-range order.\cite{Larochelle05} This lack of long-range magnetic order is commonly thought to be due to frustrated ordering of the Jahn-Teller distortions and chemical disorder of the La and Sr cations.\cite{Larochelle05,Moritomo95}  If one could directly probe the role of the dopant cation disorder in this single layered manganite, it may shed light on how these delicate low-dimensional systems organize themselves.

In this work, we synthesize random alloy and cation-ordered analogues of epitaxial La$_{2/3}$Sr$_{4/3}$MnO$_4$ films by molecular beam epitaxy (MBE). Both the random alloy and digital superlattice films exhibit the spin glass ordering temperature (20 K) seen in bulk, additional weak ferromagnetic ordering up to 130 K, and similar gapped insulator behavior at high temperatures. However, large differences in the magnetic anisotropy and the low temperature transport mechanisms suggest that ordering the dopant cations in this layered compound results in a new preferred orbital ground state.

Epitaxial La$_{2/3}$Sr$_{4/3}$MnO$_{4}$ films were grown on (001) SrTiO$_3$ (STO) substrates at 650$^{o}$C in pure ozone by MBE. Two types of epitaxial La$_{2/3}$Sr$_{4/3}$MnO$_{4}$ films were grown, random alloys and digital superlattices (SL). The random alloy films contain a random distribution of the La and Sr cations over the A-sites of the A$_2$BO$_4$ structure by co-deposition of La and Sr (1:2) for each AO rock salt monolayer. The digital SLs are cation-ordered analogues of equivalent composition, whereby complete monolayers of LaO or SrO are interleaved within the manganite A$_2$BO$_4$ structure. To achieve the composition of La$_{2/3}$Sr$_{4/3}$MnO$_{4}$, the following layering was imposed, creating a digital superlattice with a period of three A$_2$BO$_4$
units: [(SrO/MnO$_2$/LaO)(SrO/MnO$_2$/SrO)(SrO/MnO$_2$/LaO)]. All films are 33 A$_2$BO$_4$ units thick ($\sim$ 20 nm).

During growth, the film surface was monitored \emph{in-situ} by reflective high energy electron diffraction. The thickness, surface roughness and cation order of the films was investigated by x-ray reflectometry, and the crystal structure and epitaxy was probed by x-ray diffraction. Scanning transmission electron microscopy (STEM) and nanoarea electron diffraction were performed to further investigate the structure and epitaxy of the films. In-plane and out-of-plane film magnetization (M) was probed as a function of temperature (T) and applied magnetic field (H) by a Quantum Design superconducting quantum interference device (SQUID) magnetometer. Film resistivity ($\rho$) as a function of T was measured using a four-point probe geometry with a modified Quantum Design Physical Property Measurement System.

\begin{figure}
\center{\includegraphics[width=9 cm]{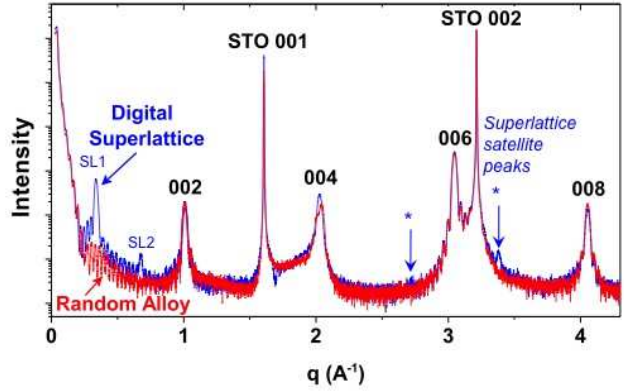}}
\caption{(Color Online.) X-ray reflectivity and diffraction of La$_{2/3}$Sr$_{4/3}$MnO$_{4}$ random alloy and digital superlattice films.}\label{XRD}
\end{figure}

X-ray diffraction of both films show strong {00l} reflections of the K$_2$NiF$_4$-type structure indicating epitaxy with the STO substrate (Fig~\ref{XRD}.) The rocking curve full-width half-max of the 006 film peaks (0.03-0.05$^{o}$) are comparable to that of the 002 STO peaks (0.01-0.03$^{o}$) demonstrating a high degree of out-of-plane crystallinity. The presence of strong SL peaks at low q corresponding to a period of 3 A$_2$BO$_4$ units confirms that the A-site cation-order imposed in the digital SL film during growth is preserved. These SL peaks would not appear if there was significant intermixing of the La and Sr cations, as evidenced by the lack of such peaks in the random alloy film. The presence of strong intensity oscillations out to high q and the clear thickness fringes around the film diffraction peaks illustrate the atomic monolayer smoothness of the film surfaces.

\begin{figure}
\center{\includegraphics[width=9
cm]{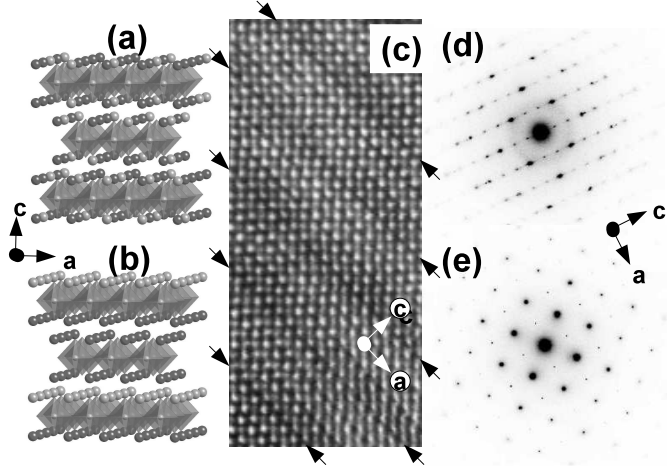}}\caption{(Color Online.) Schematic of La$_{2/3}$Sr$_{4/3}$MnO$_{4}$ as (a) random alloy and (b) digital superlattice, with La (Sr) in green (pink). (c) STEM image of digital La$_{2/3}$Sr$_{4/3}$MnO$_{4}$. Nanoarea electron diffraction of (d) film and substrate and (e) only STO substrate, for comparison.} \label{STEM}
\end{figure}

A high angle annular dark field STEM image of the digital superlattice taken along the [100] zone axis shows a high degree of crystallinity in the films, the isolated nature of the A$_2$BO$_4$ sheets typical of the K$_2$NiF$_4$-structure, and a structural expansion every third rocksalt layer (Fig~\ref{STEM}(a-c)). X-ray reflectivity simulations confirm that this periodic structural expansion is responsible for a decrease in the intensity seen in the second superlattice peak. Nanoarea electron diffraction (parallel probe size of 55 nm) of the substrate and film are shown in Fig~\ref{STEM}(d,e). A K$_2$NiF$_4$-type diffraction pattern is clearly seen showing single crystalline structure of the La$_{2/3}$Sr$_{4/3}$MnO$_{4}$ film, and alignment with the STO substrate pattern indicates in-plane registry of the film with the substrate.

\begin{figure}
\center{\includegraphics[width=9 cm]{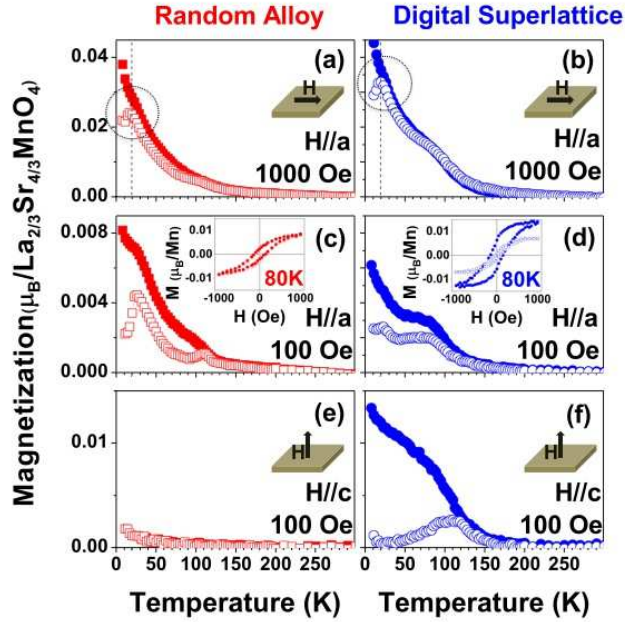}}\caption{(Color Online.) M(T) of the random alloy and digital superlattice films. Open and closed symbols were cooled in 0 Oe and H, respectively. (insets) M(H) at 80K. Open and closed symbols are H//a and H//c, respectively.} \label{Magnetism}
\end{figure}

Magnetically, both films exhibit the signature of a spin glass in-plane at 20 K, characterized by a sharp decrease of the moment for the zero-field cooled case and increase in moment for the field-cooled case (Fig~\ref{Magnetism}(a-b)).\cite{Binder86} In bulk, La$_{2-x}$Sr$_x$MnO$_4$ exhibits a spin glass (SG) phase for 0.1$<$x$<$0.6 with T$_g$ ranging from 16-25 K.\cite{Larochelle05} The digital superlattice film has individual chemical layer units of x=0 (LaSrMnO$_4$) and x=1 (Sr$_2$MnO$_4$), neither of which exhibit the spin glass phase in bulk. The fact that the observed SG phase is coincident with the random alloy film suggests that the dopant charge is not strictly tied to the chemical order of the Sr dopant cations and spreads into neighboring A$_2$BO$_4$ layers, as has also been seen in single-layered cuprates.\cite{Smadici09}

In addition to the spin glass phase, both the random alloy and digital SL films exhibit an additional magnetic transition around 130 K which is not seen in bulk La$_{2/3}$Sr$_{4/3}$MnO$_4$. Hysteresis loops show a weak ferromagnetic nature associated with this phase (Fig~\ref{Magnetism}(insets)).) While the in-plane magnetization of both types of films looks qualitatively similar (Fig~\ref{Magnetism}(c,d)), the out-of-plane magnetization differs quite dramatically (Fig~\ref{Magnetism}(e,f)). The random-alloy exhibits little to no ferromagnetic behavior out-of-plane within our resolution, and the digital SL sample exhibits a large magnetization compared to the in-plane magnetization of both films. This indicates a large difference in the magnetic anisotropy depending on cation order. Magnetic anisotropy is closely tied to spin-orbit coupling, which in the manganites is directly linked to the 3d orbitals. Thus, the significant difference in magnetic anisotropy is an indication that cation ordering may be altering the preferred orbital occupation in these layered manganites.

\begin{figure}
\center{\includegraphics[width=9 cm]{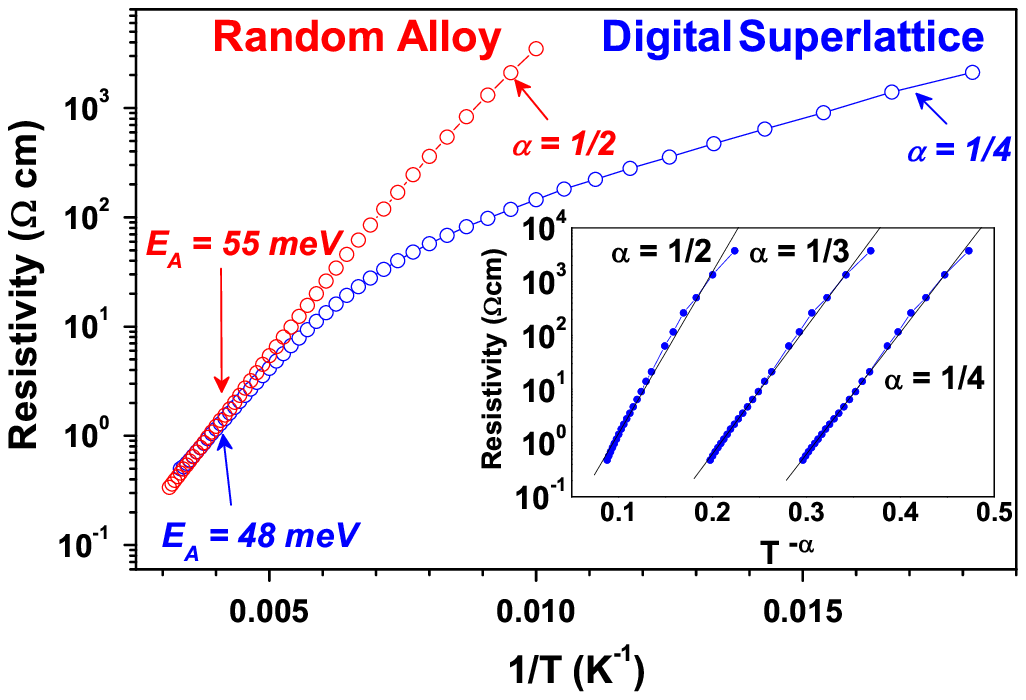}}\caption{(Color Online.) $\rho$(1/T) of the random alloy and digital superlattice films. (inset) Variable range hopping model fits to low temperature digital SL data.} \label{Transport}
\end{figure}

The resistivity of the random alloy and digital SL films shows insulating behavior for all temperatures (Fig ~\ref{Transport}.) For intermediate temperatures, both films exhibit thermally-activated transport typical of a gapped insulator with similar activation energies, E$_A$. For the random alloy E$_A$=55 meV (325K$>$T$>$145K), while for the digital SL E$_A$=48 meV (300K$>$T$>$195K). Although activation energies are unavailable for bulk La$_{2/3}$Sr$_{4/3}$MnO$_4$, Moritomo et al. found bulk LaSrMnO$_4$ (x=0) to have a E$_A$=70 meV, and that hole-doping by substitution of the La by Sr considerably reduces $\rho_{ab}$, while the system still remains insulating.\cite{Moritomo95} This is consistent with E$_A$=48-55 meV seen in our films at intermediate temperatures.

At low temperatures, the transport mechanism changes in both films, and the $\rho$(T) can be fit with a variable range hopping model. The low temperature transport of the random alloy film is best fit with $\alpha$=1/2 indicating coulomb gap behavior, consistent with strong disorder and low dimensionality. The digital SL film shows over an order of magnitude lower resistivity at 100 K than its random alloy counterpart, and is best fit with $\alpha$=1/4 indicating variable range hopping in 3D (Fig ~\ref{Transport}(inset).) Thus, although both films exhibit similar gapped insulator behavior at high temperatures, large differences in the active transport mechanism at low temperatures suggest that dopant cation order plays a pivotal role in the transport of these films. As our magnetization data indicate, ordering the cations may favor a different orbital occupancy. Depending upon the  nature of this occupancy, it would lead to changes in the bandwidth for charge transport in-plane (for example, an enhanced in-plane bandwidth would result  for d$_{x^{2}-y^{2}}$ occupancy). Thus, the measured  changes in resistivity reflect changes in both scattering and bandwidth.

We have successfully synthesized both random alloy and digital cation-ordered analogs of epitaxial La$_{2/3}$Sr$_{4/3}$MnO$_4$ films through MBE to study how cation order influences low-dimensional systems. In both films the spin glass phase found below 20K in bulk is preserved, while evidence for ferromagnetic ordering up to 130 K is seen. The magnetic anisotropy of this magnetic phase is markedly different depending on the dopant cation order. Although both films exhibit similar gapped insulator behavior at high temperatures, large differences in low temperature transport suggest that dopant cation order plays a key role in the active transport mechanism. Together, the magnetic and transport differences suggest that dopant cation order may significantly alter the preferred orbital ground state in these layered oxide systems.

This work was supported by UChicago Argonne, LLC, Operator of ANL. ANL, a U.S. DOE lab, is operated under Contract No. DE-AC02-06CH11357. B.N.C. and A.B. acknowledge support from the Digital Synthesis FWP funded by DOE-BES.  CNM was supported by the U.S. DOE-BES, under Contract No. DE-AC02-06CH11357. Electron microscopy was carried out at the Frederick Seitz Materials Research Laboratory Central Facilities, University of Illinois, which are partially supported by the U.S. DOE under Grants No. DE-FG02-07ER46453 and No. DE-FG02-07ER46471.

\end{document}